\documentclass[twocolumn, prl]{revtex4}
\usepackage{graphicx}
\pdfoutput=1
\begin{document}
\title{Spontaneous Vortex Lattices in Quasi 2D Dipolar Spinor Condensates} \author{ 
%A and B
Jian Zhang$^{\dagger \ast}$ and Tin-Lun Ho$^{\dagger}$ 
}
\affiliation{$^{\dagger}$Department of Physics, The Ohio State University, Columbus, OH 43210\\
$^{\ast}$Institute of Advanced Study, Tsinghua University, Beijing 10086, China}
\date{\today}
\begin{abstract}
Motivated by recent experiments\cite{BA}\cite{BB}, we study quasi 2D ferromagnetic condensates with various aspect ratios. We find that in zero magnetic field,  dipolar energy generates a local energy minimum with all the spins lie in  the 2D plane forming a row of {\em circular} spin textures with {\em alternating} orientation, corresponding to a packing of vortices of {\em identical}  vorticity in different spin components. In a large magnetic field, the system can fall into a long lived dynamical state consisting of an array of elliptic and hyperbolic Mermin-Ho spin textures, while the true equilibrium is an uniaxial spin density wave with a single wave-vector along the magnetic field, and a wavelength similar to the characteristic length of the long lived vortex array state. 
\end{abstract}
\maketitle

The condensates of bosons with non-zero spins, known as spinor condensate, are remarkable superfluids. 
In addition to broken gauge symmetry, they also have broken symmetries in spin space. The spin degrees of freedom lead to a variety ground states, which proliferates rapidly as the value of spin increases. Since different spin components can be mixed through spin rotation, there is considerable interplay between spin and gauge degrees of freedom, leading to a whole host of new macroscopic quantum phenomena.

 %The presence of the magnetic degrees of freedom considerably enhances the variety of macroscopic quantum phenomena. Not only does each of the $2F+1$ spin component behave like a superfluid, they can also be mixed through spin rotation. 

The simplest spinor condensates are those for spin-1 bosons, such as the $F=1$ hyperfine states of $^{23}$Na and $^{87}$Rb. The ground state of $^{23}$Na is a 
non-mangetic ``polar" condensate whereas  $^{87}$Rb is  a ferromagnetic condensate\cite{HoNaRb}. 
In the case of ferromagnetic condensates, they possess an additional ``spin-gauge" symmetry which makes non-uniform spin textures behave like vortices\cite{Jap}. The system can respond to external rotation through spin deformation. Any attempt to bend the spin will also generate vorticity. 

The magnetic nature of spinor condensates naturally leads to the consideration of dipolar energy, which is intrinsic to alkali atoms. Since dipole energy can generate non-uniform spin textures, it will generate vorticity. Indeed, Yi and Pu have shown that a $^{87}$Rb  condensate in a sufficiently flat cylindrical potential will form a circular spin texture, which is a vortex of ferromagnetic condensate with a polar core\cite{YiPu}. Recently, experiments at Berkeley have shown that a $^{87}$Rb condensate with a helical texture can decay into a random spin textures\cite{B}. By estimating the energy of the final state, the authors suggest that the phenomenon is caused by dipolar energy. 
More recently, the Berkeley group has found that a pancake like condensate of  $^{87}$Rb 
can develop a texture with periodically modulated spin-spin correlation rotating rapidly about an in-plane magnetic field\cite{BA,BB}.  They suggest that this effect is also  due to dipolar energy. 
%The effect of dipolar energy is particularly interesting for ferromagnetic condensates, as it tends to make the spins non-uniform, which in turn will generate vortices through the spin-gauge symmetry. 

Dipolar effects are highly geometry dependent.  In this paper, we would like to point out that some key features of quasi 2D $^{87}$Rb condensate due to dipolar interactions. Much of what we discuss also apply to other ferromgnetic condensates. We shall consider an anisotropic trap with frequencies $\omega_{z}>> \omega_{y}> \omega_{x}$, as in ref.\cite{B,BA,BB}. The condensate is then a thin anisotropic slab in the $xy$-plane with Thomas-Fermi radii $R_{x}, R_{y}$ such that $R_{x}/R_{y} =(\omega_{y}/\omega_{x})\equiv \lambda$, ($\lambda \sim 10$ In ref.\cite{B,BB,BA}).  In our discussions, we choose the normal to the condensate slab, $\hat{\bf z}$, to be the spin quantization axis for the condensate wavefunction $\Psi^{T}=(\psi_{1}, \psi_{0}, \psi_{-1})$, where the superscript $``{T}"$ stands for transpose. 
% is expressed along the spin quantization axis along $\hat{\bf z}$. 
%Although we focus on  $^{87}$Rb, many of these properties are general properties of  quasi 2D systems with dipolar interactions. 
We shall show that :

\noindent {\bf (1)} In zero magnetic field, dipolar energy leads to a local energy minimum consisting of a row of circular spin textures with {\em alternating} spin orientations in the long direction $x$, {\em with all the spins in the $xy$-plane}.  These textures are of the Yi-Pu type\cite{YiPu}, with a polar core and a size determined by  the Thomas-Fermi radius in the short direction, $R_{y}$. This state amounts to an array of  vortices in {\em identical vorticity} in the $\psi_{1}$ component, with $\psi_{-1}$ being its time reversed partner.  We have also found an analytic  expression that well approximates this state.

\noindent {\bf (2)} In a large magnetic field ${\bf B}$ in the $xy$-plane, the spins rotate about $\hat{\bf B}$ and experience a time averaged dipolar energy. We find that this energy is very flat in spin space around a class of textures which is an array of elliptical and hyperbolic Mermin-Ho vortices.
%, whose spins are no long confined in the $xy$-plane. 
These states are not local minima. They will eventually evolve to the true minimum, which is an 
a uniaxial spin density wave (or ``stripe phase" for short) with a single wave-vector along $\hat{\bf B}$.  Due to the flatness of the energy surface, the vortex lattice textures will be very long lived during dynamical evolutions.

%\noindent {\bf (3)} Both ${\bf (1)}$ and ${\bf (2)}$ also occur in polar condensates provided the ratio between anti-ferromagnetic spin interaction and dipolar interaction is below ??. %This shows that the system which would have been non-magnetic will distort its order parameter to generate a magnetic texture to gain dipole energy.

{\bf (A) Basic Structures:} We first consider some spin textures relevant for later discussions. We shall write the condensate wavefunction as $\Psi_{\mu} ({\bf x}) = \sqrt{n({\bf x})} \zeta_{\mu}({\bf x})$, where $n({\bf r})= \sum_{\mu}|\Psi_{\mu}|^2$ is the density, and $\zeta^{\dagger}\cdot \zeta=1$. The spin field is given by ${\bf S}= 
\Psi^{\ast}_{\mu} {\bf F}_{\mu\nu} \Psi_{\nu}= n({\bf x}) {\bf m}({\bf x})$ where ${\bf F}_{\mu\nu}$ is the spin operator, and ${\bf m}= \zeta^{\ast}_{\mu} {\bf F}^{}_{\mu\nu}\zeta^{}_{\nu}$. The general form of a 
ferromagnetic condensate is 
\begin{equation}
\zeta^{T} = e^{i\gamma} (u^2, \sqrt{2}uv, v^2), \,\,\,\,  {\rm or} \,\,\,\,\,  \frac{\sqrt{2}\psi_{1}}{\psi_{0}} = 
 \frac{\psi_{0}}{\sqrt{2}\psi_{-1}} 
 \end{equation}
where $u= e^{-i\alpha/2}{\rm cos}\beta/2, \,\,  v=e^{i\alpha/2}{\rm sin}\beta/2$. The spin is 
${\bf m}= {\rm cos}\beta \hat{\bf z} + {\rm sin}\beta({\rm cos}\alpha \hat{\bf x} + 
{\rm sin}\alpha \hat{\bf y})$, and ${\bf m}^2 =1$.  If the spin lies in the $xy$-plane, then $|u|=|v|=1/\sqrt{2}$, (or 
$|\psi_{1}|^2 : |\psi_{0}|^2 : |\psi_{-1}|^2= 1:2:1$), $\zeta$ then reduces to $\zeta^{T} = e^{i\gamma} ( e^{-i\alpha}, \sqrt{2}, e^{i\alpha})/2$. 
%\begin{equation} \zeta^{T} = e^{i\gamma} ( e^{-i\alpha}, \sqrt{2}, e^{i\alpha})/2. \end{equation}
% Let us restrict to the 2D case for the moment, and use rectangular coordinate $(x,y)$ polar coordinates $(r,\phi)$ interchangeably. 
 The following cases are of interest to us.  (Below, $(r,\phi)$ are polar coordinates.) 

\noindent 
{\em (i) Elliptic planar texture: }This corresponds to $\Psi^{T} (x,y) = \sqrt{n} ( \mp ie^{-i\phi} f(r), \sqrt{2}, \pm ie^{i\phi}f(r))$, where $f(r)$ vanishes at $r=0$ and becomes 1 beyond a healing length $\xi$. 
This describes a ferromagnet $( \mp ie^{-i\phi}, \sqrt{2}, \pm ie^{i\phi})/2$ (at large distance) with a polar core $(0,1,0)$ (at $r=0$).  All spins line up in circles in the $xy$-plane, with a  magnitude $|{\bf m}|$ shrinks from 1 to zero as $r\rightarrow 0$, (a ``meron"). This is the state found by Yi and Pu.\cite{YiPu}.
Fig.(1a) and (1b) show the spin textures of $\Psi$ with upper and lower sign, which have opposite spin orientations.  Note that reversing spin orientation does not alter the 
phase winding (or circulation) of each spin component. 

\noindent 
{\em (ii) Hyperbolic planar texture}: This is the same as (i) except with  $\phi \rightarrow -\phi$. 
See Fig.(1c) and (1d). 
%This state also has its  spins  in the $xy$-plane and a polar core. 
 
\noindent 
{\em (iii) Elliptic Mermin-Ho texture}:  This corresponds to $\Psi^{T} = \sqrt{n} ({\rm cos}^2\frac{\beta(r)}{2},  i \sqrt{2} e^{i\phi}{\rm cos}\frac{\beta(r)}{2}{\rm sin}\frac{\beta(r)}{2},  -  e^{2i\phi}{\rm sin}^2\frac{\beta(r)}{2})$ where $\beta(r)$ is an increasing function of $r$ starting with $\beta(0)=0$. As $r$ increases from 0 to a distance where $\beta=\pi/2$, $\zeta$ changes from $(1, 0, 0)$ to 
$(1, i\sqrt{2} e^{i\phi}, -e^{2i\phi})$, which has the same circular spin texture as $(i)$ at large distance.  It
differs from (i) in that it is everywhere ferromagnetic. Vortex singularities can be eliminated without damaging the ferromagnetic order. See Fig.(1e).

\noindent 
{\em (iv)  Hyperbolic Mermin-Ho texture}: This is the same as $(iii)$ with $\phi\rightarrow - \phi$. 
%and is the non-singular analog of (ii). See figure (1f). 

\begin{figure}
\includegraphics[width=2.8in]{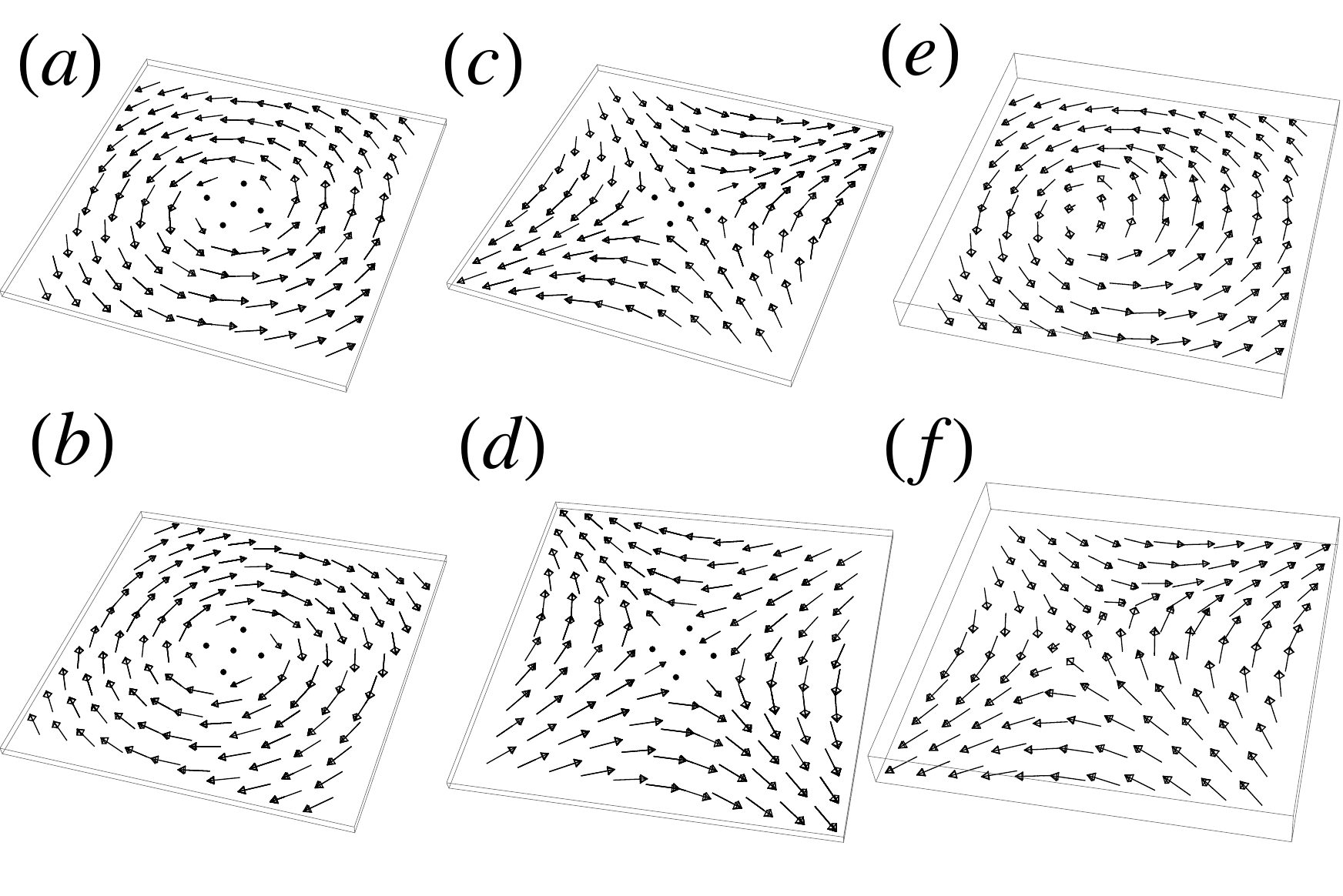}
\caption{(1a) and (1b) are elliptic planar textures with identical vorticity in $\psi_1$ and $\psi_{-1}$. 
(1c) and (1d) are hyperbolic planar spin textures with identical vorticity. All these spin textures (1a)-(1d) have a polar core. (1e) and (1f) are elliptic and hyperbolic Mermin-Ho textures. The cores are fully ferromagnetic. }
\end{figure}

{\bf (B) Energetic considerations:} We shall discuss the zero field case at zero temperature.  Although $B\neq 0$ in current experiments, we first discuss this case because of its fundamental importance.
%So long as there is no structural phase transition before temperature reaches $T_{c}$, the spin texture at $T=0$ should be the same as those at finite temperatures. 
In zero magnetic field, the energy is $E[\Psi] =  {\cal T} + {\cal V} + {\cal U} + {\cal V}_{D}$, where ${\cal T}=  \int \frac{\hbar^2}{2M}\nabla |\Psi_{\mu}|^2$ is the kinetic energy, ${\cal V}= \int V({\bf x}) n({\bf x})$ is  harmonic trap, $V({\bf x}) = \frac{1}{2} M (\omega_x^2 x^2 + \omega_y^2 y^2 + \omega_z^2 z^2) $,  ${\cal U}= \int  \frac{1}{2}[ c_{o} n^2 + c_{2} {\bf S}^2]$ describes density repulsion $(c_{0}>0)$ and ferromagnetic interaction  $(c_{2}<0)$\cite{HoNaRb}, ${\cal V}_{D} = \frac{1}{2} g_{D} \int S_{i}({\bf x})D_{ij}({\bf x}-{\bf x'})S_{j}({\bf x'})$ is the dipolar energy,  where $D_{ij}({\bf x}) = \delta_{ij} /x^3 - 3x_{i} x_{j}/x^5$, $g_{D}= \gamma^2$, $\gamma= g_{F}\mu_{B}$, $g_{F}=-1/2$ is the magnetic $g$-factor, and $\mu_{B}$ is the Bohr magneton.  Noting that $D_{ij}({\bf x})= - \nabla_{i} \nabla_{j}
\frac{1}{|{\bf x}|} - \frac{4\pi}{3}\delta_{ij} \delta({\bf x})$,  we have 
\begin{equation}
{\cal V}_{D} = \frac{g_{D}}{2} \left[ \int_{\bf x,x'} \frac{ Q({\bf x}) Q({\bf x'})}{|{\bf x} - {\bf x'}|}  - \frac{4\pi}{3}\int_{\bf x}{\bf S}^2({\bf x}) \right] , \,\,\,\,\,
\label{VD} \end{equation}
where $Q({\bf x}) = \nabla\cdot {\bf S}({\bf x})$. The first term in eq.(\ref{VD}) is positive definite, the optimum spin configuration is the one that satisfies $\nabla \cdot {\bf S}=0$ while keeping $|{\bf m}|=1$ to gain maximum ferromagnetic energy. 
 The difficulty in finding the equilibrium textures is that these two conditions are not always compatible. 
 
For $^{87}$Rb, $c_{2}/c_{0}= 0.005$\cite{BA}, the healing length for spin is much longer than that for density, and exceeds the thickness of the  ``pancake"  condensates. The spin degrees of freedom is effectively 2D, in the sense that 
% spin structure $\zeta$ for a flat ``pancake"  condensate  is then independent of $z$, effectively 2D. We then have 
\begin{equation}
\Psi_{\mu}(x,y,z) = \sqrt{n(x,y,z)}\zeta_{\mu}(x,y),
\label{form} \end{equation}
and  ${\bf m}$ depends only on $(x,y)$. %In the experiments  in ref.(\ref{B2}), even though the system is thin slab, the system is still in 3D regime where the density is give by the Thomas Fermi for can still be described by Thomas
Since $n({\bf x})$ is  mirror symmetric about the $xy$-plane , the term $QQ$ in eq.(\ref{VD}) can be replaced by 
%$ [ s_{z}({\bf r}_{\perp}) \partial_{z}n({\bf r}))  [s_{z}({\bf r'}_{\perp})][ \partial_{z}n({\bf r'})] + 
$Q_{//}({\bf x}) Q_{//}({\bf x'})  + Q_{\perp}({\bf x}) Q_{\perp}({\bf x'})$, where $Q_{//}({\bf x})= m_{z}({\bf x}_{\perp})  \partial_{z}n({\bf x})$, and $Q_{\perp}({\bf x})= \nabla_{\perp}\cdot {\bf S}_{\perp}({\bf x})$, $\nabla_{\perp} = (\nabla_{x}, \nabla_{y})$, ${\bf S}_{\perp} = (S_{x}, S_{y})$. The energy is minimized by $m_{z}=Q_{\perp}=0$,  or simply $\nabla_{\perp}\cdot {\bf S} =0$ with
${\bf S}$ in the $xy$-plane. In regions where density is uniform, these conditions can be satisfied by the elliptic planar texture $(i)$ mentioned above. Indeed, circular spin alignment are prevalent in all cases we studied. Their presence, however, often require some fractions of their hyperbolic counterparts to facilitate their close packing, even though the latter  are not as energetically favorable.  

%Another point to stress is that dipolar energy scales as $n^2$, whereas kinetic energy scales as $n$. As a result, the spin texture is mainly determined by the dipole energy, even the the resulting structure may be costly in gradient energy 

{\bf (C) Our calculation}: %To speed up our calculation in the effectively 2D limit, 
%While one can perform a full numerical minimization to find the stationary point of $E[\Psi]$, we shall go to the quasi 2D limit which will allow us to speed up our calculation considerably. We 
For simplicity, 
we take the density to be a Gaussian along $z$ and Thomas-Fermi in $xy$-plane,  $n(x,y, z) = w(z)^2 n_{FT}(x,y)$, where $w(z)= 
e^{-z^2/2d^2}/(\pi^{1/4}d^{1/2})$, $d$ is the width of the condensate along $z$, and $n_{TF}({\bf r})= [\mu- \frac{1}{2}(M\omega^{2}_{x}x^2 + M\omega^{2}_{y}y^2)]/\tilde{c}_{o}$,  $\tilde{c}_{o} = c_{o}/(\sqrt{2\pi} d)$, ${\bf r} =(x,y)$, and $\mu$ is the chemical potential determined by the total number of particles. We have ignored the effect of $c_{2}$ on the density since  $c_{2}/c_{0}= 0.005$\cite{BA}. 
With this density, all the variational variables are contained in $\zeta$, 
%(Although the density profile in the flat condensate in ref.(\cite{B2}) along $z$ is Thomas-Fermi rather than Gaussian, the spin texture should not change significantly by going to the quasi 2D limit.)   The only variable is the normalized complex vector $\zeta$, 
which can be parametrized by five real fields,  $\{ X_{i}({\bf r}), i=1..5\}$.  
%The energy $E[\Psi]$ is then  a functional of these five fields. 
To locate the energy minima, we evolve the variables $X_{i}({\bf r})$ by a dissipative dynamics  $dX_{i}({\bf r}, t)/dt = - \Gamma_{i} \frac{ \delta E[X_{j}]}{\delta X_{i}({\bf r}, t)}$, where $\Gamma_{i}>0$.  Since $dE/dt = - \int {\rm d}{\bf r} \Gamma_{i} (\delta E/\delta X_{i}({\bf r}, t))^2 <0$, this evolution forces the energy to decrease, coming to a stop only when a local minimum ($\delta E/\delta X_{i}=0$) is reached.
Starting this evolution with different initial conditions, one can locate the energy minimum while mapping  out the energy surface in the space of $\zeta({\bf r}) $, or $\{ X_{i}({\bf r}) \}$\cite{details}. 

{\bf (D) Zero field case}: %The whole host of dipolar effects in this regime can be detected in future experiments. 
We have performed imaginary time evolution  for a system with $10^6$ particles  for different aspect ratios $\lambda$ and different initial states: 
%in zero magnetic field with $10^6$ particles and different aspect ratios $\lambda$. These initial states include  
random configurations, spiral spin textures, and vortex lattice textures. In all cases, the final state is either a uniform texture, or a state $\Psi^{(o)}$ with a row of circular planar texture (type  (i) in Section $({\bf A})$) with  alternating orientations. The texture for the $\lambda=10$ case is shown fig.2.  All the spins indeed lie in the $xy$-plane as discussed in Section ${\bf (B)}$. 
%The spin-spin correlation function of this state $G ({\bf r})$ as defined in ref.\cite{B2}  is shown in fig.2b. 

Based on the discussions in $({\bf A})$, we find that our numerical result $\Psi^{(o)}$ can be described accurately by   \begin{equation}
\tilde{\Psi} (x,y,z)^{T}=  \sqrt{n}(e^{i \Phi({\bf r})} f({\bf r})  , \sqrt{2}, e^{-i \Phi({\bf r})} f({\bf r}))
\label{tildePsi} \end{equation}
where $n=w^2(z) n_{TF}({\bf r})$, ${\bf r}=(x,y)$, $\Phi({\bf r})$ is the phase angle of the $1D$ vortex lattice, 
\begin{equation}
\Phi(x,y) = \pi/2 + {\rm arg}  {\cal P}(z), \,\,\,\,\, z=x+iy,
\label{Phi} \end{equation}
 and ${\cal P}(z) = \prod_{n} (z - a -nb)$, $n = 0, \pm 1, \pm 2, .., \pm M$ is a polynomial describing an array of vortices of {\em identical} circulation centered at $x=a$,  separated by distance $b$ along the $x$ axis. We find that when $\lambda$ is an odd (even) integer,  $a=0$ ($a=b/2$). 
 %For condensates with odd (even) aspect ratios, $\lambda = 1,3,.. $, ($\lambda = 2,4,.. )$, we have $a=0$ ($a=b/2$). 
$M$ is the smallest integer such that $(M+1)b> R_{x}$.  
The function $f$ describes the vortex core with size $\xi$\cite{xxx} . 
%%%, which we assume to be $f({\bf r})= {\rm tanh}^2(r/\xi)$\cite{comment}.   The angle $\pi/2$ in eq.(\ref{Phi}) is crucial for creating a circular spin texture. The opposite orientation of the neighboring circular  units in see Fig. 2 is due to the change of sign of ${\cal P}(z)$ as one passes through a vortex from left to right. 
%alternating circular spin texture.  %which is made up of circular planar textures with alternating orientation. 
%The alternating  orientation is caused by the alternating sign of ${\cal P}(z)$ as one passes through the sequence of vortices from left to right along the $x$-axis. 
For all aspect ratios we examined, the optimal value of $b$ (the size of the circular planar unit)  
and $\xi$ (core size ) is found to be  $\sim 1.5$ and $\sim0.25$ in units of $R_{y}$ (the shorter Thomas-Fermi in the $xy$-plane. 

%By varying $b$ and $\xi$, we find that the difference $\int |\Psi^{(o)}- \tilde{\Psi}|^2/ \int |\Psi^{(o)}|^2$ can be minimized to $2\%$.  The optimal value for $b$ (which is the size of the circular planar unit) and core size $\xi$ is found to be  $\sim 1.5$ and $\sim0.25$ in units of $R_{y}$ (the shorter Thomas-Fermi in the $xy$-plane) for all aspect ratios we examined. 
The packing of vortices with identical vorticity in both $\tilde{\psi}_{1}$ and $\tilde{\psi}_{-1}$ may seem surprising, for it  costs kinetic energy. The reason is that this is the only packing that leads to circular spin textures, which is strong favored by dipolar energy. With dipolar energy scales as $n^2$ and kinetic energy scales as $n$, the condensate wavefunction will be determined by the former. 
%In Fig.3, we compare the energies of the uniform texture state and and the vortex row state for different aspect ratios for a system of $10^{6}$ particles\cite{details}. 
Our calculations show that the vortex row state $\Psi^{(o)}$ is the ground state for $\lambda \leq 2 $. For $\lambda>2$, the ground state is a uniform spin texture along $\hat{\bf x}$, while $\Psi^{(o)}$ is a local minimum\cite{zzz}. 

\begin{figure}
\includegraphics[width=2.8in]{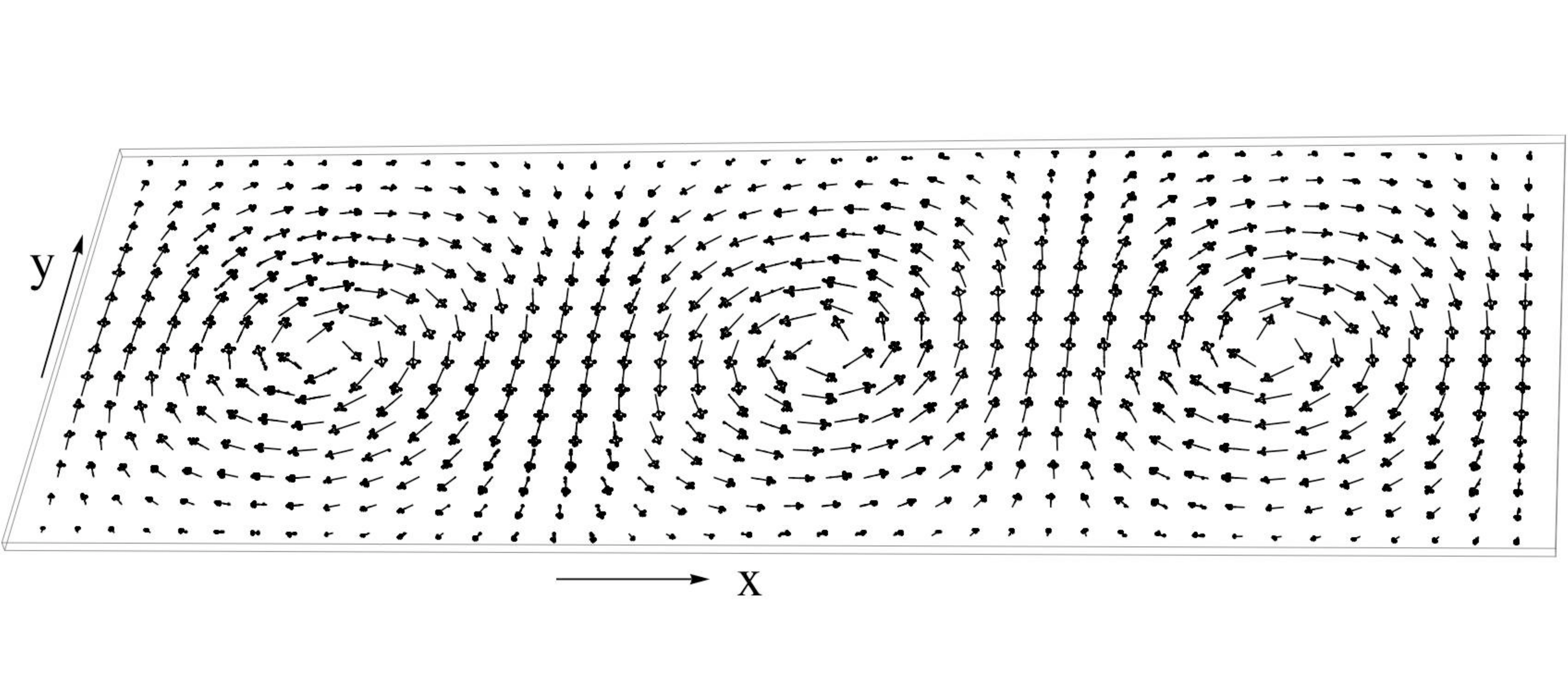}
\caption{ The middle section of the texture of an energy minimum of a system with aspect ratio $\lambda=10$ in zero magnetic field. The state consists of 
a row of elliptic planar textures  with alternating spin orientations. All spins lies in the $xy$-plane. See Fig.(1a) and (1b).}
\end{figure}

%In the case of random initial condition, we also observed an interesting phenomenon .....

{\bf (E) A large magnetic field ${\bf B}$}:  In a magnetic field ${\bf B}$, the energy acquires linear and quadratic Zeeman shift, ${\cal E}_{1}=  - \int \gamma {\bf B}\cdot  \Psi^{\ast}_{\mu}{\bf F}_{\mu\nu}\Psi_{\nu}$ and 
${\cal E}_{2} = \int q \Psi^{\ast}_{\mu}(\hat{\bf B}\cdot {\bf F})^2_{\mu\nu}\Psi_{\nu}$. For sufficiently large magnetic field, the spin texture of the system will rotate about ${\bf B}$. 
%To make contact with experiments\cite{B2}, we  are interested in the spin textures that rotates about $\hat{\bf B}$.   
 In the frame rotating about $\hat{\bf B}$ with Lamour frequency, the linear Zeeman shift is transformed away. Dipolar energy, however, acquires time dependent terms in this frame that oscillates with the Lamour frequency.  Over times much longer than the Lamour cycle, the system sees an averaged dipolar energy 
 $\overline{{\cal V}'_{D}} = \int S_{i}({\bf r})\overline{D}_{i,j}({\bf r}-{\bf r'})S_{j}({\bf r'})$, where
 \begin{equation}
 \overline{D}_{i,j} ({\bf x}) =  
 \frac{ ( 3  \hat{\bf B}_{i} \hat{\bf B}_{j} - \delta_{ij})}{2|{\bf x}|^3} \left(    1 - \frac{3(\hat{\bf B}\cdot {\bf x})^2}{|{\bf x}|^2}\right). 
% \left( 3  \hat{\bf B}_{i} \hat{\bf B}_{j} - \delta_{ij}\right)
 \end{equation}
The total time averaged energy is then $\overline{E} = {\cal T} + {\cal V} + {\cal U} + \overline{{\cal V}_{D}} + {\cal E}_2$. 
 Note that $\overline{E}$ conserves  $\hat {\bf B} \cdot \int {\bf S}$. 
In ref.\cite{BA,BB}, $\hat{\bf B}$ is aligned with $\hat{\bf x}$ up to a few degrees.  The observed state with 
periodic spin texture is found to have $\int S_{x}\sim 0$ and with most spins lie perpendicular to $\hat{\bf x}$.

To search for the stationary states of $\overline{E}$ with $\int S_{x}\sim 0$, we performed the 
dissipative dynamics mentioned in Section ${\bf (C)}$ with a great variety of initial conditions.  Depending on the quadratic Zeeman energy $(q)$, two types of stationary state emerge. For $q/\hbar\omega_{y} > (q/\hbar\omega_{y} )_{c}\sim 0.1$,  the equilibrium state is 
%essentially described by  $\zeta = (- {\rm cos} \frac{K\hat{\bf B}\cdot {\bf r}}{2}, 0,  {\rm sin} \frac{K\hat{\bf B}\cdot {\bf r}}{2} )$, which 
is an uniaxial spin density wave with  ${\bf S}({\bf r}) = \hat{\bf z } \cos (K\hat{\bf B}\cdot{\bf r})$.
%$\zeta = (- {\rm cos} \frac{K\hat{\bf B}\cdot {\bf r}}{2}, 0,  {\rm sin} \frac{K\hat{\bf B}\cdot {\bf r}}{2} )$, which gives ${\bf S} = \hat{\bf z} 
For the parameters used in footnote \cite{details}, we find a wavelength $2\pi/K\sim 12.5 a_{y}\sim 25 \mu m$,  ($a_{y}= \sqrt{\hbar/(M\omega_{y})}$) and is independent of the angle between $\hat{\bf  B}$ and $\hat{\bf x}$. 
%where $m({\bf r})$ is an oscillatory function with wavevector ${\bf K}$ along $\hat{\bf B}$, and with the spins (which rotates about $\hat{\bf B}$) are all along a direction $\hat{\ell}$ perpendicular to $\hat{\bf B}$.  We find  that $K \approx 1/\xi_{D}$, where $\xi_{D}$ is the dipole length defined as  $\hbar^2/(2M \xi_{D}^2) = g_{D}\overline{n}$, and $\overline{n}$ is the average density.  Using the numbers in  \cite{details}, $\xi_{D}\approx 3.0 \mu m$. 
For $q/\hbar\omega_{y}  < (q/\hbar\omega_{y} )_{c}$, the system will settle in a state consisting of two large uniform spin domains along $\hat{\bf x}$ and $-\hat{\bf x}$, with a pair of elliptic and hyperbolic Mermin-Ho texture sandwiched in between. 

In additional to these two equilibrium states, we also find a class of spin textures which are essentially  distorted vortex lattices which remain in the imaginary time evolution for a very long time,  reflecting the flatness of the energy surface 
%(or almost vanishing  $\partial \overline{E}/\partial \zeta$ 
in the neighborhood of these states. The underlying structure of these distorted lattice are pairs of elliptic and hyperbolic MH vortices,  (See Fig.4), with essentially zero spin projection along $\hat{\bf x}$, $\int S_{x}/N\sim 10^{-3}$. These states remain long lived for $q/\hbar\omega_{y}  < 0.2$ and begin to evolve more rapidly toward the stripe phase as $q$ increases.  We have calculated the spin-spin correlation function $G({\bf r})$ as defined in ref.\cite{BA} for the vortex arrays in Fig.4. The result is shown in Fig. 5 where bright color represents high value of $G({\bf r})$. The underlying lattice structure of the disordered spin texture Fig.4  shows up a lattice of bright spots in Fig. 5. 

In Figure 6, we have plotted $|S({\bf K})|^2$, where ${\bf S}({\bf K})$ is the Fourier transform of spin texture ${\bf S}({\bf r})$ shown in Figure 4. The almost periodic structure in real space shows up as intense spots in $K-$space.  Taking the brightest spot closest to the origin on the right, we find $(a_{y}/\lambda_{x}, a_{y}/\lambda_{y})  \sim (0, 0.08)$, or  $\sqrt{\lambda_{x}^2 + \lambda_{y}^2} \sim 25 \mu m$,  comparable to the period found in the stripe phase, and is the same order of the ``dipolar length"\cite{details}.  We note that the corresponding plot in ref.\cite{BA} shows a length scale $\sim 10 \mu m$, differing from our result by a factor of 2.  This 
difference  may be due to our Gaussian approximation of the actual density normal to the plane, which will contribute to systematic errors in the energies of all textures. However, such systematic error will not change the fact that  both the stripe phase and the spin lattice textures have similar length scales -- a property that can be verified by experiments.

%ref.\cite{B2}. 
 
 %, and is again of the order of the dipole length $\xi_{D}$.  

%The average separation between vortices is about 30$\mu m$ with the parameters in \cite{details}, which is similar to the spatial  period of the stripe phase. 

% structure similar to that in ref.\cite{B2}.  The lattice constant 

%The lattice spacing is found to be of comparable to $\xi_{D}$. 

% We also find that for these  spin vortex lattice states, the ratio between longitudinal and transverse component of the spin, defined as $\int S_{x}({\bf r})^2/ \int ( S_{y}^2({\bf r})  + S_{z}^{2}({\bf r})  )$ decreases as $q$ increases.  At $q/\hbar\omega_{y} =0.15$, this ratio is $\sim 1/2$. 
%In Figure 6, we have plotted $|S({\bf K})|^2$, where ${\bf S}({\bf K})$ is the Fourier transform of spin texture ${\bf S}({\bf r})$ shown Figure 4. Periodic structure in real space now shows up intense spots in $K-$space.  The length scale extracted from the most intense spot for our example\cite{details} is ??? $\mu$m, comparable with that found in ref.\cite{B2}, and is again of the order of the dipole length $\xi_{D}$.  % These findings are consistent with those in ref.\cite{B2}. 

\begin{figure}
\includegraphics[width=2.8in]{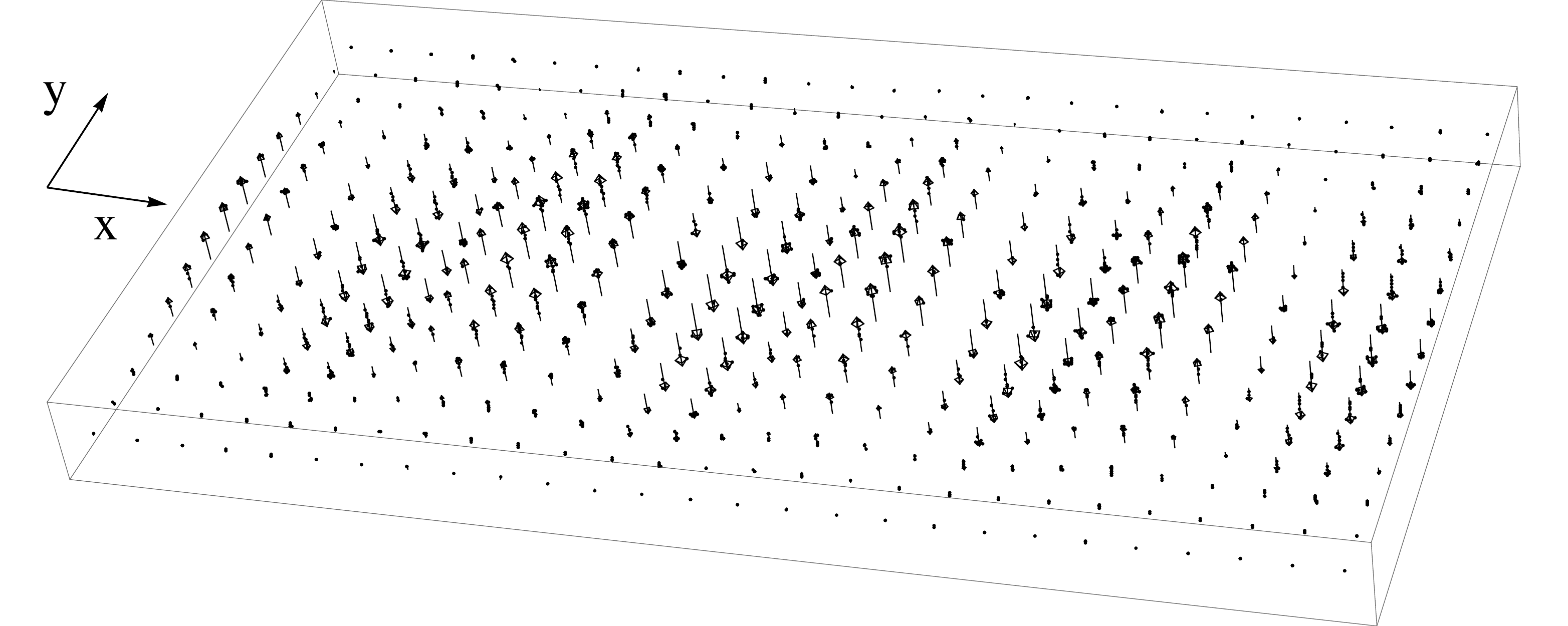}
\caption{The middle section of the texture of the equilibrium state of $\overline{E}$ for a system with aspect ratio $\lambda=10$.}
\end{figure}

\begin{figure}
\includegraphics[width=2.8in]{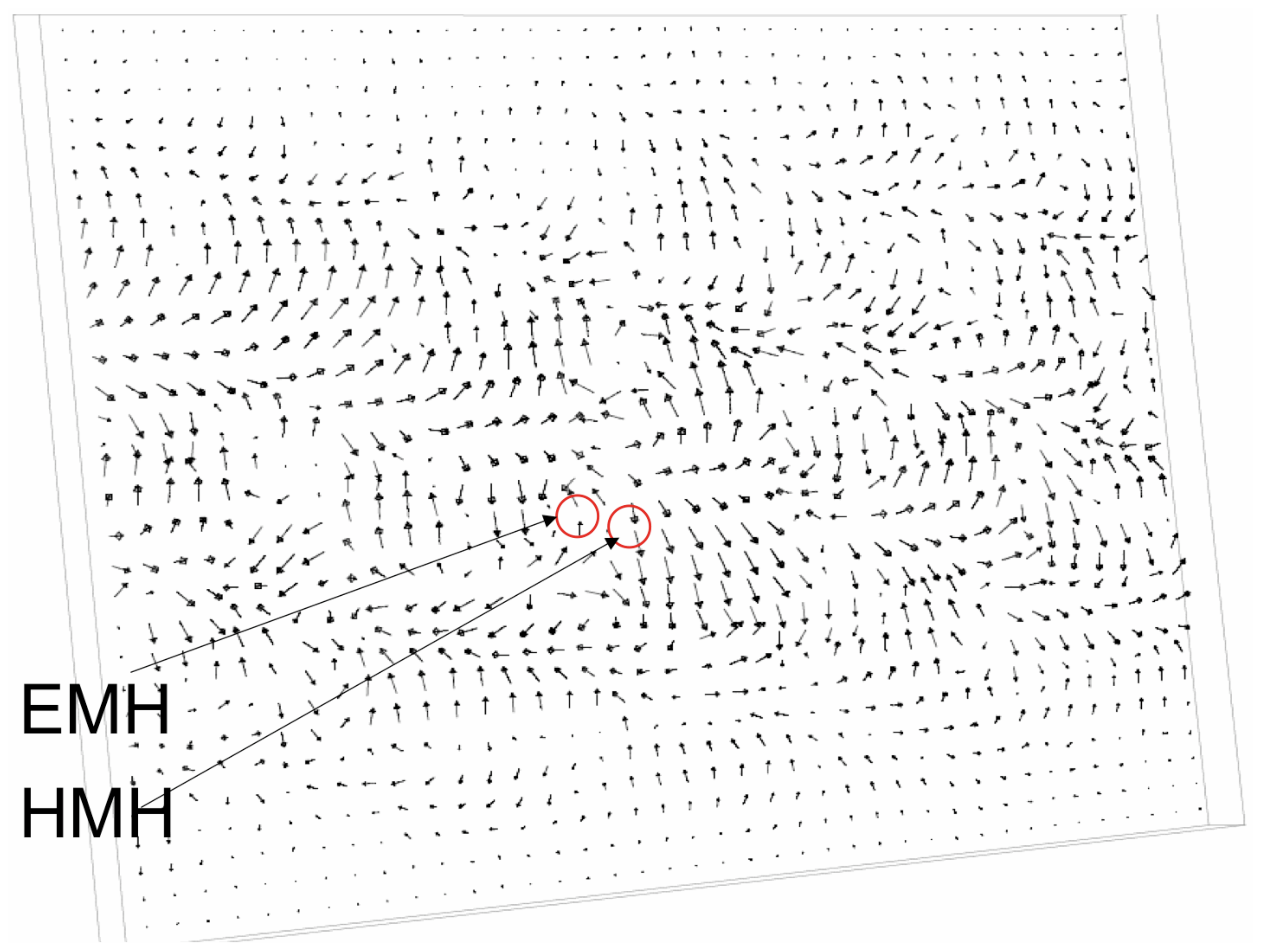}
\caption{A long lived state of the time average energy $\overline{E}$ -- an array of elliptic and hyperbolic Mermin-Ho vortex pair. }
\end{figure}

\begin{figure}
\includegraphics[width=2.8in]{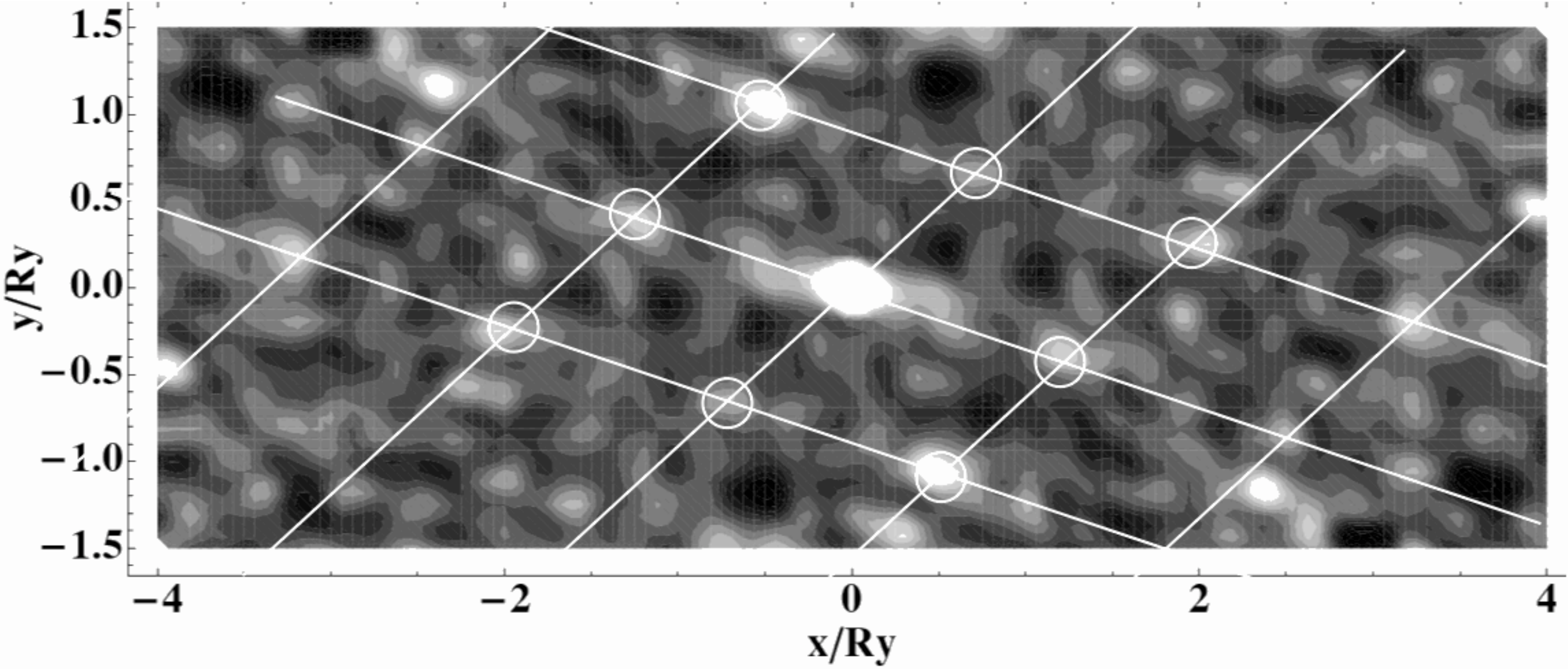}
\caption{spin-spin correlation of the state in Fig.4. Note that the bright spots form an almost square lattice.}
\end{figure}

\begin{figure}
\includegraphics[width=1.5in]{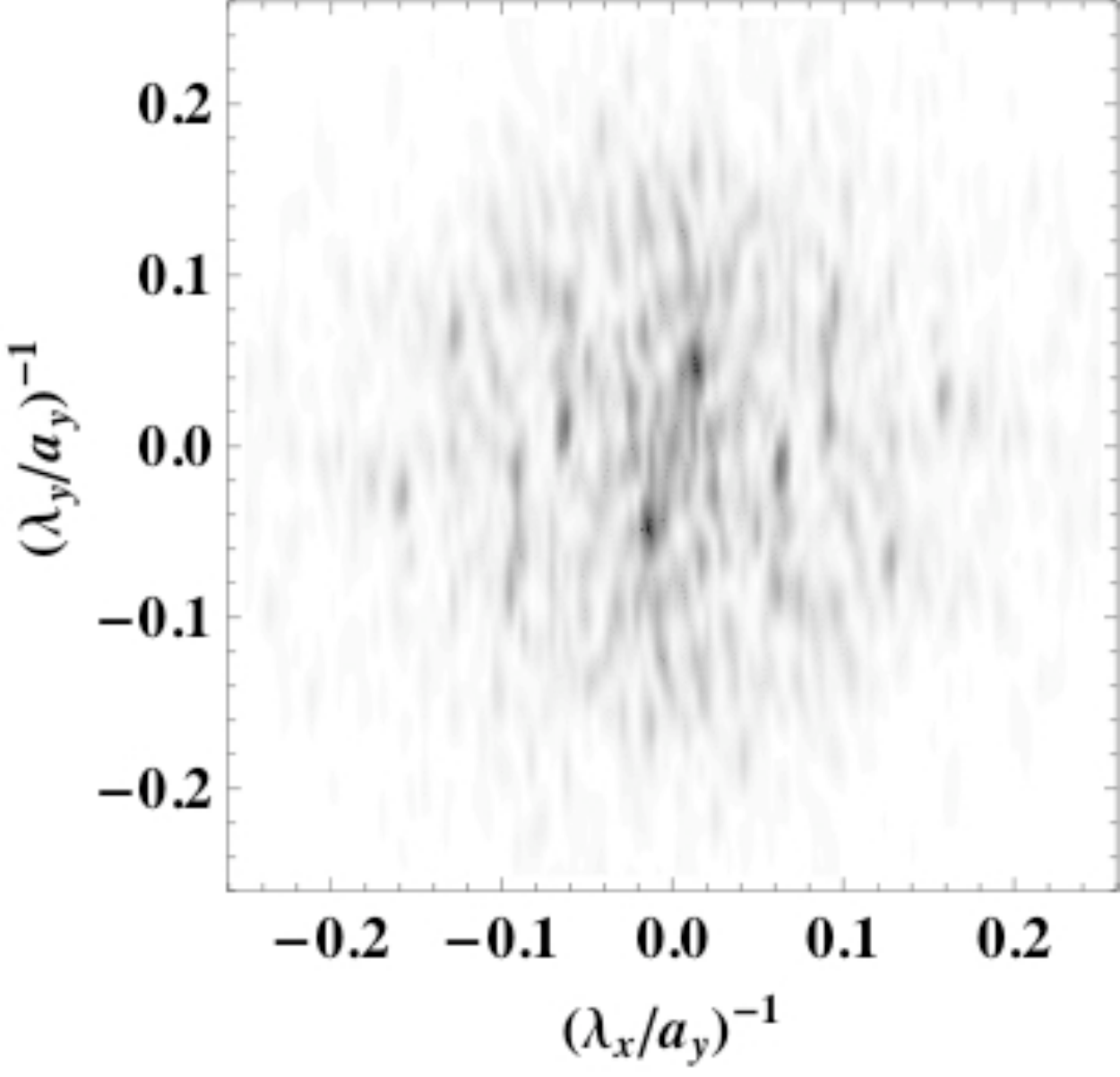}
\caption{$|{\bf S}({\bf K})|^2$ of the state in Fig.4;  ${\bf K} = 2\pi(\lambda_{x}^{-1}, \lambda_{y}^{-1} )$.}
\end{figure}

Since the long lifetime in this dissipative evolution is due to the almost vanishing $\delta E/\delta \zeta$, similar long lifetime will also occur in real time evolution as it is driven by the same derivative. 
Our findings, which is consistent with those in  ref.(\cite{BA,BB}),  suggest that the observed periodic structure is a long lived dynamical state.  They also predict two  equilibrium configurations:  a row of alternating circular spin texture in zero field, and a single wavevector spin stripe phase in large magnetic field.  Our method is also applicable to other dipolar condensates .

We thank Dan Stamper-Kurn and especially Mukund Vengalattore for discussions. This work is supported by ARO Grant W911NF0710576 for the DARPA OLE Program, and by NSF Grants PHY-05555576, DMR0705989.

\vspace{0.2in}

% To find the energy minimum, we consider an imaginary time evolution of $\zeta$ according to the equation:  \begin{equation} \frac{d \zeta}{d t} = - c \frac{d E[\zeta, \zeta^{\ast}]}{d \zeta^{\ast}}  \end{equation}


\begin{thebibliography}{99}
\bibitem{zzz} The energy of the vortex row state $\Psi^{(o)}$ is basically a sum of the ``self" energy of individual circular spin textures. Their interaction energy can be ignored because  the effective ``dipole moment" of each unit is essentially zero. 
On the other hand, a  uniform texture can be viewed as an 1D array  of  ``giant dipoles" obtained by 
 turning each  circular spin unit in the  original circular spin array into a unit uniform texture. 
 While each of these ``dipole" has higher self energy than the circular unit, they interact much strongly with each other. As $\lambda$  increases, the interaction energy in the uniform spin units will outweigh the self energy of the circular units. The switch appears at $\lambda =2$.
\end{thebibliography}
\end{document}